\documentclass[12pt,twoside,a4paper,amsmath,amssymb,aps,showkeys,showpacs]{revtex4}

\usepackage{amsfonts}
\usepackage{graphics}
\usepackage{epsfig}
\usepackage{fancyheadings}

\begin{document}
\DeclareGraphicsExtensions{.jpg,.pdf,.mps,.png,}

\thispagestyle{myheadings}

\rhead[]{}
\lhead[]{}
\chead[V.L.Kalashnikov]{Chirped Dissipative Solitons}

\title{Chirped Dissipative Solitons}

\author{Vladimir L. Kalashnikov}
\email{kalashnikov@tuwien.ac.at}
\affiliation{Institut f\"{u}r Photonik, TU Wien, Gusshausstrasse 27/387, A-1040 Vienna, Austria}

\begin{abstract}
The analytical theory of chirped dissipative soliton solutions of nonlinear complex Ginzburg-Landau equation is exposed. Obtained approximate solutions are easily traceable within an extremely broad range of the equation parameters and allow a clear physical interpretation as a representation of the strongly chirped pulses in mode-locked both solid-state and fiber oscillators. Scaling properties of such pulses demonstrate a feasibility of sub-mJ pulse generation in the continuous-wave mode-locking regime directly from an oscillator operating at the MHz repetition rate.
\end{abstract}


\maketitle

\section{Introduction}
\label{introduction}
The nonlinear complex Ginzburg-Landau equation (CGLE) has such a wide horizon of application
that the concept of "the world of the Ginzburg-Landau equation" has become broadly established \cite{kramer}. The CGLE is used in quantum optics, modeling of Bose-Einstein condensation, condensate-matter physics, study of non-equilibrium phenomena,
and nonlinear dynamics, quantum mechanics of self-organizing dissipative systems, and
quantum field theory. In particular, this equation being a generalized form of the master mode-locking
equation \cite{kaertner,akh1} provides an adequate description of pulse localized inside a laser oscillator or propagating along a
nonlinear fiber.

The exact analytical solutions of CGLE are known only for a few of cases, when they represent the dissipative solitons and
some algebraic relations on the parameters of equation are imposed \cite{akh1}. More general solutions can be revealed on
basis of the algebraic nonperturbative techniques \cite{conte}, which, nevertheless, are not developed sufficiently still.

The perturbative methods allow obtaining the dissipative soliton solutions in the vicinity of the Schr\"{o}dinger solitonic sector \cite{agrawal} of the CGLE \cite{malomed}. Another approximate method utilizes the reduction of infinite-dimensional space of CGLE to finite-dimensional one on basis of the method of moments \cite{akh2}.

In this article I present the direct approximate integration technique for the generalized CGLE \cite{kalash1,kalash2}. This technique is suitable to  a physically important sector, which is represented by the chirped solitary pulse
solutions, or the chirped dissipative solitons (CDSs) of the CGLE. The method under consideration does not impose any functional constraints
on the solution and equation parameters. Additionally, the dimensionality of the obtained CDS is reduced in comparison
with the parametric space of the CGLE. As a result, the CDS characteristics become easily traceable.

As was found \cite{kalash3}, the CDS is energy scalable due to strong pulse chirp, which results in the substantial soliton stretching. Such a stretching reduces the CDS peak power and, thereby, leads to the soliton stabilization. Simultaneously, a strong chirp provides a sufficiently broad spectrum allowing the pulse compression by the factor 10$\div$100. As a result, one can obtain the femtosecond pulses with over-$\mu$J energy at the fundamental MHz repetition rate of an oscillator. This allows the table-top high-intensity experiments with the level of intensity of the order of 10$^{14}$ W/cm$^2$ \cite{morgner}. Thus, an analysis of the CSP solutions of nonlinear CGLE is of interest not only from the theoretical
but also from the practical points of view.

\section{Approximate integration of the CGLE}
\label{integration}
Let the 1+1-dimensional generalized CGLE be written down in the following form \cite{kalash1,kalash2}

\begin{eqnarray}\label{GL}
\begin{gathered}
 \frac{\partial }{{\partial z}}a\left( {z,t} \right) =  - \sigma a\left( {z,t} \right) + \alpha \frac{{\partial ^2 }}{{\partial t^2 }}a\left( {z,t} \right) + \Sigma \left[ {\left| {a\left( {z,t} \right)} \right|^2 } \right]a\left( {z,t} \right) +  \\
 i\beta \frac{{\partial ^2 }}{{\partial t^2 }}a\left( {z,t} \right) - i\left( {\gamma  + \chi \left| {a\left( {z,t} \right)} \right|^2 } \right)\left| {a\left( {z,t} \right)} \right|^2 a\left( {z,t} \right),
  \end{gathered}
\end{eqnarray}
\noindent where the field envelope $a(z,t)$ depends on the propagation (longitudinal) coordinate $z$ and on  the local time $t$ (``transverse'' coordinate; for an interpretation of (\ref{GL}) in the terms of Bose-Einstein condensate see, for instance, \cite{pitaev,konotop}). The real part of the right-hand side of (\ref{GL}) (nonlinear Ginzburg-Landau equation) describes the dissipative effects: i) net-loss with the coefficient $\sigma$, which is the difference between unsaturated net-loss and saturated gain; ii) spectral dissipation with the squared inverse bandwidth coefficient $\alpha$, and iii) nonlinear gain, which is some function $\Sigma$ of the power $|a|^2$. The imaginary part of the right-hand side of (\ref{GL}) (nonlinear Schr\"{o}dinger equation) describes the nondissipative effects: i) group-delay dispersion (GDD) with the coefficient $\beta$, and ii) self-phase modulation (SPM) with the coefficient $\gamma$. The SPM can be saturable or enhanceable with the corresponding coefficient $\chi$.

The nonlinear gain (or self-amplitude modulation (SAM)) in Eq. (\ref{GL}) describes the dependence of dissipation on the field. To provide a soliton formation, $\Sigma$ has to grow with power at least for a small $P(z,t)\equiv |a(z,t)|^2$. In an oscillator, the SAM can be provided by the different mechanisms: self-focusing inside an active medium, nonlinear polarization dynamics, bleaching of an absorber (e.g., a semiconductor quantum-well structure), etc. (see \cite{kaertner,r}). Below two main functional forms of $\Sigma$ will be considered: i) cubic-quintic SAM term $\Sigma \equiv \kappa P - \zeta \kappa P^2$, and ii) perfectly saturable SAM $\mu \kappa P/(1+\kappa P)$. Here $\kappa$ is the effective inverse loss saturation power, $\zeta$ is the parameter of the SAM saturation ($\zeta >$0 to provide the stability against collapse), and $\mu$ is the SAM depth. The first type of SAM describes a Kerr-lens mode-locked oscillator, and the second type corresponds to an oscillator mode-locked by a semiconductor saturable absorber. It should be noted, that the functional form of $\Sigma$ is not decisive for the method under consideration.

The main assumption underlying our method is that the dissipative soliton has a large chirp. Mainly, this assumption is valid in the positive dispersion range ($\beta>$0) of Eq. (\ref{GL}), but the is the CDS also in the negative, or anomalous, dispersion range ($\beta<$0, see \cite{akh3}). As a result of large chirp, the CDS is temporally stretched that admits the adiabatic approximation ($T \gg \sqrt{\beta}$, $T$ is the soliton width), and the fast variation of soliton phase admits the stationary phase method in the Fourier domain \cite{kalash4,kalash5}. Formally, the approximations under consideration mean that $\beta \gg \alpha$ and $\gamma \gg \kappa$ (or $\gamma \gg \mu \kappa$). That is the nondissipative effects prevail over the dissipative ones.

\subsection{Cubic-quintic CGLE}
Let the SAM term in Eq. (\ref{GL}) be written in the form

\begin{equation}\label{cubic-quintic}
    \Sigma \left[ {\left| {a\left( {z,t} \right)} \right|^2 } \right] = \kappa P(z,t) - \zeta \kappa P(z,t)^2,
\end{equation}
\noindent and the desired traveling wave solution of (\ref{GL}) be
\begin{equation}\label{trav}
a\left( {z,t} \right) = \sqrt {P\left( t \right)} \exp \left[ {i\phi \left( t \right) - iqz} \right].
\end{equation}
\noindent Here $\phi(t)$ is the phase and $q$ is the soliton wavenumber. Substitution of (\ref{cubic-quintic}) and (\ref{trav}) in (\ref{GL}) with subsequent application of the adiabatic approximation results in \cite{kalash2,kalash6,kalash7}

\begin{eqnarray}\label{adiab}
\begin{gathered}
 \beta \Omega ^2  = q - \gamma P - \chi P^2 , \\
 \beta \left( {\frac{\Omega }{P}\frac{{dP}}{{dt}} + \frac{{d\Omega }}{{dt}}} \right) = \kappa P\left( {1 - \zeta P} \right) - \sigma  - \alpha \Omega ^2 ,
 \end{gathered}
 \end{eqnarray}
\noindent where $\Omega \equiv d\phi(t)/dt$ is the deviation of instant frequency from the carrier one. The latter is assumed to coincide with the minimum of spectral dissipation. The solution of (\ref{adiab}) is

\begin{equation}\label{p}
    P = {{\left( {\sqrt {\gamma ^2  + 4q\chi  - 4\chi \beta \Omega ^2 }  - \gamma } \right)} \mathord{\left/
 {\vphantom {{\left( {\sqrt {\gamma ^2  + 4q\chi  - 4\chi \beta \Omega ^2 }  - \gamma } \right)} {2\chi ,}}} \right.
 \kern-\nulldelimiterspace} {2\chi}},
\end{equation}
\noindent which exists in the limit of $\chi\rightarrow$0:

\begin{equation}\label{limitp}
P = {{\left( {q - \beta \Omega ^2 } \right)} \mathord{\left/
 {\vphantom {{\left( {q - \beta \Omega ^2 } \right)} \gamma }} \right.
 \kern-\nulldelimiterspace} \gamma }.
\end{equation}

The condition of power positivity requires an existence of maximum frequency deviation $\Delta ^2  = {q \mathord{\left/
 {\vphantom {q \beta }} \right.
 \kern-\nulldelimiterspace} \beta }$.

Second equation in (\ref{adiab}) leads to

\begin{eqnarray} \label{omega}
\begin{gathered}
  \frac{{d\Omega }}
{{dt}} = \frac{{\sigma  + \alpha \Omega ^2  - \frac{{\kappa \left(
{A - \gamma } \right)}} {{4\chi ^2 }}\left( {2\chi  + \zeta \gamma
- \zeta A} \right)}}
{{\beta \left[ {4\chi \beta \Omega ^2  - \left( {A - \gamma } \right)A} \right]}}\left( {A - \gamma } \right)A, \hfill \\
  A = \sqrt {\gamma ^2  + 4\beta \chi \left( {\Delta ^2  - \Omega ^2 } \right)} . \hfill \\
\end{gathered}
\end{eqnarray}

In the limit of $\chi \rightarrow$0, one has \cite{kalash5,kalash8}

\begin{eqnarray}\label{rq-c}
\begin{gathered}
 \gamma P\left( 0 \right) = \beta \Delta ^2  = \frac{{3\gamma }}{{4\zeta }}\left( {1 - \frac{c}{2} \pm \sqrt {\left( {1 - \frac{c}{2}} \right)^2  - \frac{{4\sigma \zeta }}{\kappa }} } \right), \\
 \frac{{d\Omega }}{{dt}} = \frac{{\beta \zeta \kappa }}{{3\gamma ^2 }}\left( {\Delta ^2  - \Omega ^2 } \right)\left( {\Omega ^2  + \Xi ^2 } \right), \\
 \beta \Xi ^2  = \frac{\gamma }{\zeta }\left( {1 + c} \right) - \frac{5}{3}\gamma P\left( 0 \right), \\
 \end{gathered}
 \end{eqnarray}
\noindent where $c=\alpha \gamma/\beta \kappa$ and the equation for the peak power $P(0)$ (or, equally, for $\Delta$) results from the regularity condition ${{d\Omega } \mathord{\left/
 {\vphantom {{d\Omega } {dt <  \pm \infty }}} \right.
 \kern-\nulldelimiterspace} {dt <  \pm \infty }}$. Such a condition for Eq. (\ref{omega}) leads to ($\varsigma \equiv \sigma \zeta/\kappa$, $\eta \equiv \zeta \gamma/\chi$):

\begin{eqnarray}\label{delta-full}
\begin{gathered}
\Delta ^2  = \frac{\gamma } {{16\zeta \beta \left( {\frac{c} {\eta} +
1} \right)}}\times \\
\left[ {\frac{{2\left( {3 + \frac{c} {\eta} + \frac{4} {\eta}}
\right)\left( {2 + \frac{c} {2} + \frac{{3\eta}} {2} \pm \sqrt {\left(
{c - 2} \right)^2  - 16 \varsigma \left( {1 + \frac{c} {\eta}} \right)} }
\right)}} {{1 + \frac{c} {\eta}}} - 12 - 3c - 9\eta - \frac{{32\varsigma }} {{
\eta}}} \right].
\end{gathered}
\end{eqnarray}

In the cubic nonlinear limit of Eqs. (\ref{GL},\ref{cubic-quintic}) ($\chi \rightarrow$0, $\zeta \rightarrow$0), which admits an exact CDS solution \cite{haus}, one has

\begin{eqnarray}\label{schroed}
\begin{gathered}
 a\left( {z,t} \right) = \sqrt {P\left( 0 \right)} {\mathop{\rm sech}\nolimits} \left( {\frac{t}{T}} \right)^{1 - i\psi } e^{ - iqz} , \\
 \alpha \Delta ^2  = \frac{{3\sigma c}}{{2 - c}},\,\,\gamma P\left( 0 \right) = \beta \Delta ^2 , \\
 \psi  = \frac{{3\gamma }}{{\kappa \left( {1 + c} \right)}},\,\,T = \frac{{3\gamma }}{{\kappa \Delta \left( {1 + c} \right)}}. \\
 \end{gathered}
 \end{eqnarray}

One may see that the maximum parametric dimension of (\ref{GL},\ref{cubic-quintic}) is 4, but the parametric dimension of the approximate solutions is reduced (the maximum dimension corresponds to (\ref{p},\ref{omega}) and equals to 3, i.e. $\varsigma$, $\eta$ and $c$).

Next assumption allows further simplification. Since the phase $\phi(t)$ is a rapidly varying function due to a large chirp, one may apply the method of stationary phase to the Fourier image $e(\omega)$ of $a(t)$ \cite{kalash4}. Then, the spectral profile corresponding to (\ref{p},\ref{omega}) can be written as

\begin{equation}
p\left( \omega \right) \equiv |e\left( \omega \right) |^{2} \approx
\frac{\pi \left( A-1\right) \left( \left( A-1\right) c \eta+4\left(
2\omega ^{2}-\Delta ^{2}\right) \right) \rm H\left( \Delta ^{2}-\omega
^{2}\right) }{cA\left( \left( A-1\right) \left( c\left(
\varsigma+\eta+\eta^{2}+\omega ^{2}\right) +\eta\left( \Delta ^{2}-\omega ^{2}\right)
\right) -2\left( \eta+1\right) \left( \Delta ^{2}-\omega ^{2}\right)
\right) }, \label{spectrum}
\end{equation}

\noindent where the normalizations: $ t' = t\left(
{{\kappa  \mathord{\left/
 {\vphantom {\kappa  \zeta }} \right.
 \kern-\nulldelimiterspace} \zeta }} \right)\sqrt {{\kappa  \mathord{\left/
 {\vphantom {\kappa  {\alpha \zeta }}} \right.
 \kern-\nulldelimiterspace} {\alpha \zeta }}}
$, $ \Delta '^2  = {{\Delta ^2 \alpha \zeta } \mathord{\left/
 {\vphantom {{\Delta ^2 \alpha \zeta } \kappa }} \right.
 \kern-\nulldelimiterspace} \kappa }
$, $ \Omega '^2  = {{\Omega ^2 \alpha \zeta } \mathord{\left/
 {\vphantom {{\Omega ^2 \alpha \zeta } \kappa }} \right.
 \kern-\nulldelimiterspace} \kappa }
$, $ P' = \zeta P$, and $ E' =
E\left( {{\kappa  \mathord{\left/
 {\vphantom {\kappa  \gamma }} \right.
 \kern-\nulldelimiterspace} \gamma }} \right)\sqrt {{{\kappa \zeta } \mathord{\left/
 {\vphantom {{\kappa \zeta } \alpha }} \right.
 \kern-\nulldelimiterspace} \alpha }}
$ are used (primes are omitted). $
A \equiv \sqrt {1 + {{4\left( {\Delta ^2  - \Omega ^2 } \right)} \mathord{\left/
 {\vphantom {{4\left( {\Delta ^2  - \Omega ^2 } \right)} {\eta c}}} \right.
 \kern-\nulldelimiterspace} {\eta c}}}
$ and
 $\rm H$ is the Heaviside's function.

\begin{figure}[h!]
\includegraphics[width=7.3cm]{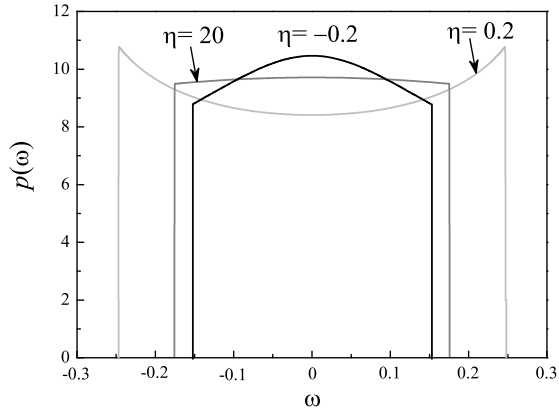}
\caption{\label{fig1} Spectra corresponding to the $-$ branch (i.e. $-$ sign in Eq. (\ref{delta-full})) of CDS for $c=$1, $\varsigma=$0.01 and different values of $\eta$. $\eta>$0 corresponds to an enhancement of the SPM with the power growth and $|\eta|\rightarrow \infty$ corresponds to an absence of the quintic SPM, i.e. $|\chi| \rightarrow$0.}
\end{figure}

As it has been demonstrated in \cite{kalash2}, the truncated at $\pm \Delta$ spectra (\ref{spectrum}) have convex, concave, and concave-convex vertexes (Fig. \ref{fig1}). When $\chi \rightarrow$0, only convex truncated spectra remain \cite{kalash5}:

\begin{equation}\label{redspectrum}
    p\left( \omega  \right) \approx \frac{{6\pi \gamma }}{{\zeta \kappa }}\frac{{{\mathop{\rm H}\nolimits} \left( {\Delta ^2  - \omega ^2 } \right)}}{{\Xi ^2  + \omega ^2 }}.
\end{equation}

At last, for the cubic nonlinear limit of Eq. (\ref{GL}) ($\chi \rightarrow$0, $\zeta \rightarrow$0) one may obtain

\begin{equation}\label{cubic}
    p\left( \omega  \right) \approx \frac{{6\pi \beta }}{{\kappa \left( {1 + c} \right)}}{\mathop{\rm H}\nolimits} \left( {\Delta ^2  - \omega ^2 } \right).
\end{equation}

\noindent That is the truncated flat-top spectrum, which approximates the spectrum of exact solution of the cubic CGLE (\ref{GL}). However, the exact spectrum is expressed through the beta-function that troubles its analytical study.

An appearance of the concave spectrum (\ref{redspectrum}) from the flat-top one (\ref{cubic}) can be illustrated by means of the perturbation analysis in the spectral domain. The approximate method under consideration allows easily performing such an analysis. In the case of the cubic CGLE with the real quintic term as a perturbation, the equation for the Fourier image of the soliton perturbation $\tilde{f}\left(\omega\right)$ can be written as \cite{akh1}

\begin{equation}
\left[k\left(\omega\right)-q\right]\tilde{f}\left(\omega\right)+\frac{1}
{\pi}\intop_{-\infty}^{\infty}d\omega^{'}U\left(\omega-\omega^{'}\right)
\tilde{f}\left(\omega^{'}\right)+\frac{1}{2\pi}\intop_{-\infty}^{\infty}
d\omega^{'}V\left(\omega-\omega^{'}\right)\tilde{f}^{*}\left(\omega^{'}\right)=
S\left(\omega\right),\label{eq:eqpqrturbation}
\end{equation}
\noindent where $
k\left( \omega  \right) =  - \beta \omega ^2  - i\left( {\sigma  + \alpha \omega ^2 } \right)
$ is the complex wavenumber of linear wave, the kernels $U$ and $V$ are the Fourier images of $|a(t)|^2$ and $a(t)^2$, respectively ($a(t)$ corresponds to the time-dependent part of solution (\ref{schroed}) and co-propagation of a stable CDS with a sustained perturbation is assumed). The assumption, that the perturbation is in-phase with the CDS allows the replacement $V\otimes \tilde{f}^{*}  \leftrightarrow U\otimes \tilde{f}$ in (\ref{eq:eqpqrturbation}). It should be noted, that such an assumption is only conjecture for the CDS (compare with \cite{kalash9}). The Fourier image $S(\omega)$ of the perturbation source can be obtained from the perturbation term $-\kappa \zeta P(t)^2 a(t)$ by means of the method of stationary phase (see above):

\begin{equation}\label{source}
S\left( \omega  \right) =  - \frac{{i\zeta \beta ^2 }}{{\gamma ^2 }}\sqrt {\frac{{6\pi \beta \kappa }}{{\left( {1 + c} \right)}}} \left( {\Delta ^2  - \omega ^2 } \right)^2 \exp \left[ {\frac{{3i\gamma \omega ^2 }}{{2\kappa \left( {\left( {\Delta ^2  - \omega ^2 } \right)\left( {1 + c} \right)} \right)}}} \right]{\mathop{\rm H}\nolimits} \left( {\Delta ^2  - \omega ^2 } \right).
\end{equation}

For the kernel $U$, one has

\begin{equation}\label{U}
   U\left( {\omega  - \omega '} \right) =  - \frac{{9\pi \gamma \beta \left( {\gamma  + i\kappa } \right)}}{{2\kappa ^2 \left( {1 + c} \right)^2 }}\left( {\omega  - \omega '} \right){\mathop{\rm csch}\nolimits} \left[ {\frac{{3\gamma \pi \left( {\omega  - \omega '} \right)}}{{2\kappa \Delta \left( {1 + c} \right)}}} \right].
\end{equation}

Resulting equation is the Fredholm equation of second kind and can be solved by the Neumann series method so that the
iterative solution is

\begin{equation}\label{neumann}
   \tilde f_n \left( \omega  \right) = \frac{{S\left( \omega  \right)}}{{k\left( \omega  \right) - q}} - \frac{3}{{2\pi \left( {k\left( \omega  \right) - q} \right)}}\int\limits_{ - \infty }^\infty  {d\omega 'U\left( {\omega  - \omega '} \right)} \tilde f_{n - 1} \left( {\omega '} \right),
\end{equation}
\noindent where $
\tilde f_n \left( \omega  \right)
$ is the $n$-th iteration and $
\tilde f_0 \left( \omega  \right) = {{S\left( \omega  \right)} \mathord{\left/
 {\vphantom {{S\left( \omega  \right)} {\left( {k\left( \omega  \right) - q} \right)}}} \right.
 \kern-\nulldelimiterspace} {\left( {k\left( \omega  \right) - q} \right)}}
$.

\begin{figure}[h!]
\includegraphics[width=7.3cm]{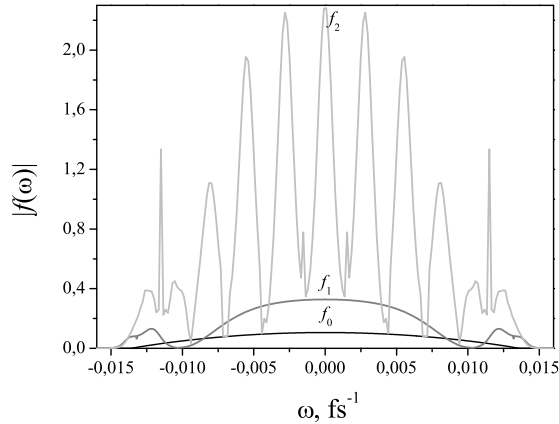}
\caption{\label{fig2} Absolute values of the successive terms in the Neumann series (\ref{neumann}) for the real quintic nonlinear perturbation (tildes are omitted). $\alpha=$16 fs$^2$, $\beta=$250 fs$^2$, $\kappa=0.04\gamma$, $\sigma=2.5\times10^{-4}$, and $\zeta=0.2\gamma$.}
\end{figure}

The successive terms of (\ref{neumann}) are shown in Fig. \ref{fig2}. One can see that the tabletop spectrum becomes convex and some oscillating substructure develops. The last phenomenon explains the results of \cite{kalash10}, where the cause of both regular and chaotic pulsations of CDS has been attributed to an excitation of the solitonic internal modes revealed by oscillating structure in Fig. \ref{fig2}.

It is convenient to represent the CDS existence regions in the form of the
so-called master diagrams \cite{kalash5}. The coordinates of such diagram are the normalized energy and the $c$-parameter. The energy $
E \equiv \int_{ - \Delta }^\Delta  {\frac{{d\omega }}{{2\pi }}} p\left( \omega  \right)
$ can be easily obtained from Eqs. (\ref{spectrum},\ref{redspectrum},\ref{cubic}) and connected with the $\sigma$-parameter through $
\sigma  \approx \delta \left( {{E \mathord{\left/
 {\vphantom {E {E^*  - 1}}} \right.
 \kern-\nulldelimiterspace} {E^*  - 1}}} \right)
$, where $
\delta  \equiv \left. {{{d\sigma } \mathord{\left/
 {\vphantom {{d\sigma } {dE}}} \right.
 \kern-\nulldelimiterspace} {dE}}} \right|_{E = E^* }
$ and $E^*$ is the energy of continuous-wave solution of Eq. (\ref{GL}) \cite{kalash5}.

\begin{figure}[h!]
\includegraphics[width=7.3cm]{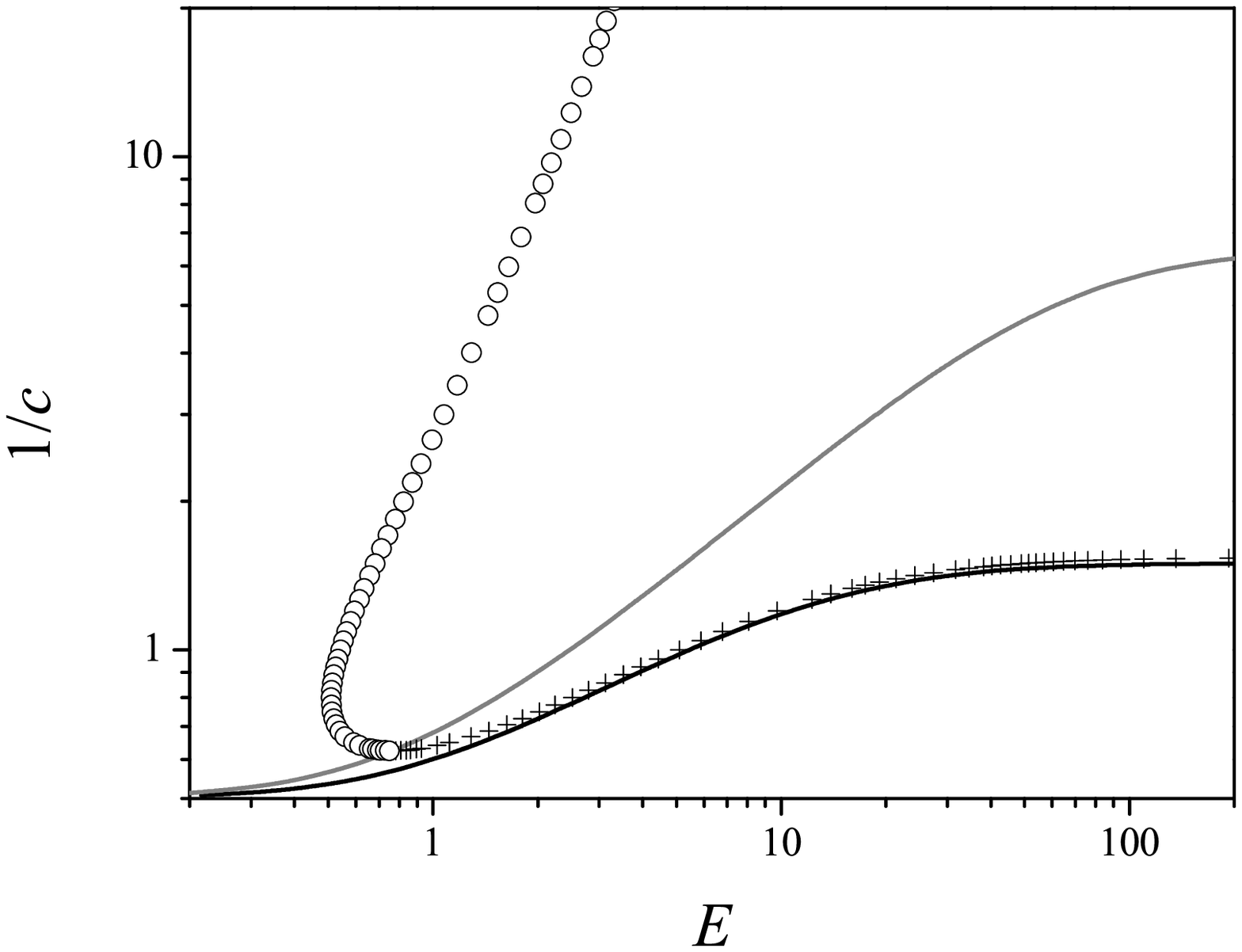}
\caption{\label{fig3} The master diagram for the CDS of Eqs. (\ref{GL},\ref{cubic-quintic}), $\chi=$0. There exists no
CDS below the black solid line (here $\sigma<$0, i.e. the vacuum of (\ref{GL}) is unstable). Black solid curve corresponds to $\varsigma=$0. Gray curve divides the regions, where the $+$ and $-$
branches of (\ref{delta-full}) exist. Crosses (circles) correspond to the $+$($-$) branch for
$\varsigma$=0.01.}
\end{figure}

The master diagram for the CDS is shown in Fig. \ref{fig3} for the case of vanishing quintic SPM ($\chi=$0). The CDS is two-parametric in this case. The solid
curve shows the border of the CDS existence $\varsigma=$0. Below this border, the vacuum of Eq. (\ref{GL}) is unstable. The gray curve divides the existence regions for the $+$ and $-$ branches (the corresponding signs in Eq. (\ref{delta-full})). The branches merge along this curve. Crosses (circles) represent the curve along which there exists the +(-) branch for some fixed value of $\varsigma$ (isogain curve).

The master diagram reveals four significant differences
between the $+$ and $-$ branches of CDS. The first one is that the $-$ branch has
lower energy than the $+$ branch for a fixed $1/c$. The second
difference is that the $+$ branch of an isogain has a finite bound for $E\rightarrow \infty$. In this sense, the $+$ branch is energy
scalable, that is its energy growth does not require a substantial change of $c$. The $-$ branch is not energy scalable, that is
the energy growth along this branch needs a substantial decrease of $c$ (e.g.,
owing to a dispersion growth). The third difference is that the $+$ branch verges on $\varsigma$=0 within a whole range of $E$. The fourth difference is that the $-$ branch has a Schr\"{o}dinger limit $\zeta,\chi \rightarrow$0 (Eq. (\ref{schroed},\ref{cubic})).

\begin{figure}[h!]
\includegraphics[width=7.3cm]{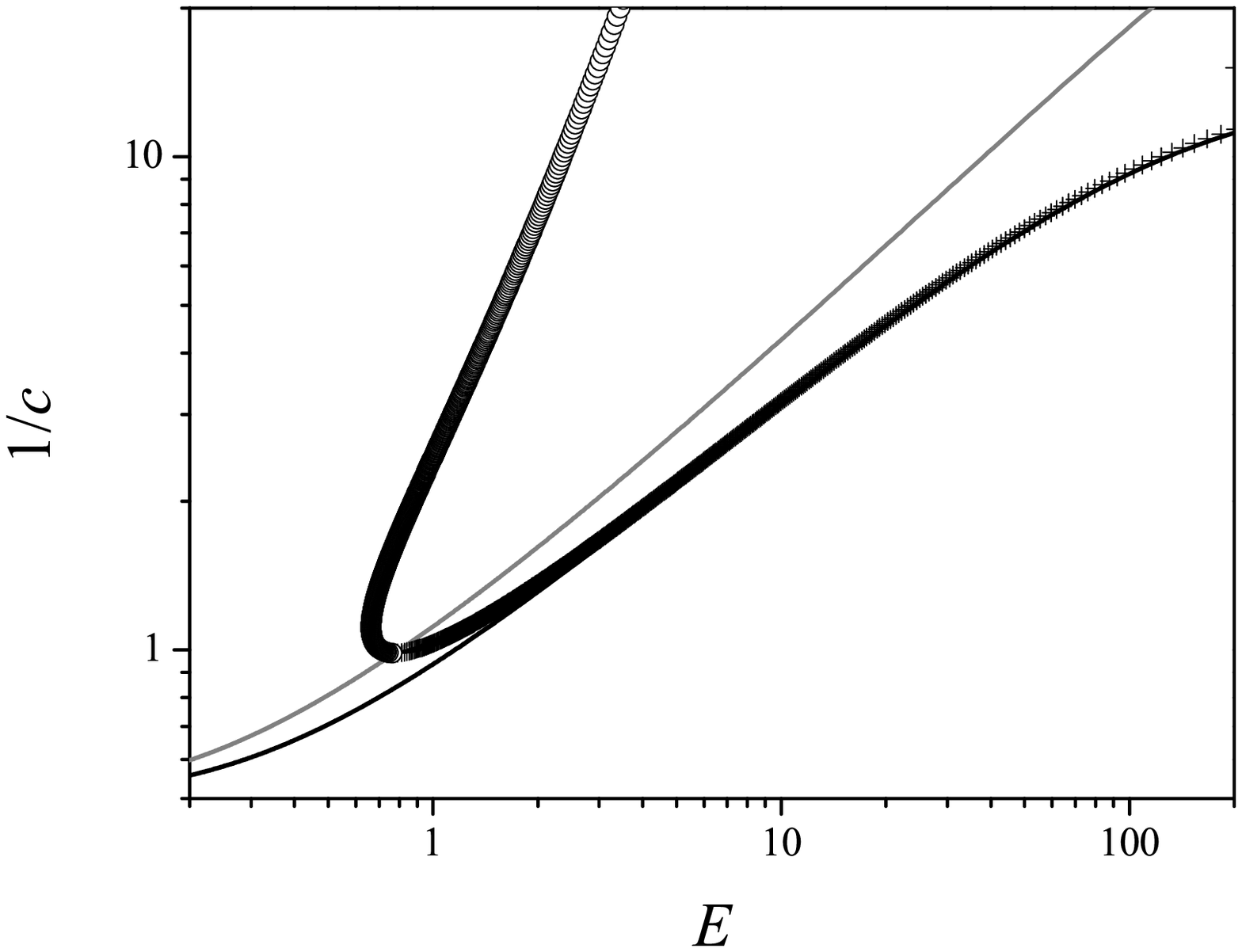}
\caption{\label{fig4} The master diagram for the CDS of Eqs. (\ref{GL},\ref{cubic-quintic}), $\eta=$0.2. There exists no
CDS below the black solid line (here $\sigma<$0, i.e. the vacuum of (\ref{GL}) is unstable). Black solid curve corresponds to $\varsigma=$0. Gray curve divides the regions, where the $+$ and $-$
branches of (\ref{delta-full}) exist. Crosses (circles) correspond to the $+$($-$) branch for
$\varsigma$=0.01.}
\end{figure}

For $\chi \neq$0, the CDS is three-parametric. Contribution of the positive quintic SPM (i.e., $\chi>$0 and the SPM enhances with power) narrows
the CDS existence region (Fig. \ref{fig4}). This means that smaller $c$ is
required to provide the CDS existence for a fixed $E$. That is,
since the positive quintic SPM means an enhancement of the
SPM with power, a SPM enhancement has to be compensated,
for instance, by a dispersion increase $c\propto1/\beta$. One
can see from Fig. \ref{fig4}, that the $+$ branch region narrows substantially with a growth of positive $\chi$ within a whole range of $E$. However, it is important to note, that the range of $c$, where the CDS with a fixed $\varsigma$
exists, is defined by the difference between i) the point of
intersection of the isogain with the boundary between the $+$
and $-$ branches and ii) the isogain asymptotic for $E\rightarrow \infty$. As
a result, the range of $c$, where some isogain exists, broadens with $\chi>$0 (Fig. \ref{fig4}).

It is of interest to compare the described structure of a master diagram with the results presented in Refs. \cite{akh4,akh5}, where the concept of ``dissipative soliton resonance'' has been raised. It is clear from the above, that this concept is interrelated with our concept of ``energy scalable'' CDS. That is a dissipative soliton resonance corresponds to an isogain asymptote of $E\rightarrow \infty$ in Figs. \ref{fig3},\ref{fig4} (see also \cite{kalash2,kalash5}). For instance, the approach of \cite{akh4} based on the method of moments predicts the resonance condition $1/c\approx3.16$ for $\sigma=$0 and $\chi=$0 versus $1/c \approx$1.5 in Fig. \ref{fig3}. This quantitative disagreement disappears, when the resonance condition is found on basis of the numerical simulations \cite{akh5} instead of the approximate method of \cite{akh4}. In agreement with \cite{akh4,akh5}, the asymptotic value of $1/c$ increases with $\chi>$0 (Fig. \ref{fig4} and \cite{kalash2}) and decreases with $\chi<$0 \cite{kalash2}. One can conclude, that our approximate approach is more general and accurate than that of Ref. \cite{akh4}. Moreover, the energy scalable CDSs found on basis of our approach are easily traceable because they are lying on three-dimensional surfaces of reduced parametrical space of the CGLE.

\subsection{CGLE with a perfectly saturable SAM}
Such a type of SAM corresponds to a semiconductor saturable absorber, which is extensively used for obtaining the CDS in both solid-state and fiber oscillators \cite{naumov,wise,morgner}. If the CSP width (usually, few picoseconds) excesses the SAM relaxation time $T_r$ (hundreds
of femtoseconds), the SAM function can be written in the form
\begin{equation}\label{sesam}
    \Sigma \left[ {\left| {a\left( {z,t} \right)} \right|^2 } \right] =
\frac{{\mu \kappa P\left( {z,t} \right)}}{{1 + \kappa P\left( {z,t} \right)}},
\end{equation}
\noindent where $
\kappa  = {{T_r } \mathord{\left/
 {\vphantom {{T_r } {E_s S}}} \right.
 \kern-\nulldelimiterspace} {E_s S}}
$ is the inverse saturation power, $E_s$ is the saturation energy fluency, $S$ is the mode area. $\chi=$0 will be assumed below.

The above considered method results in \cite{kalash1,kalash6,kalash11}

\begin{eqnarray}\label{sesam2}
\begin{gathered}
 \kappa P\left( 0 \right) = \frac{\alpha }{{\mu c}}\Delta ^2  = \frac{3}{{4c}}\left[ {2\left( {1 - \varsigma } \right) - c \pm \sqrt \Upsilon  } \right], \\
 \frac{{d\Omega }}{{dt}} = \frac{\alpha }{{3\beta }}\frac{{\left( {\Delta ^2  - \Omega ^2 } \right)\left( {\Xi ^2  - \Omega ^2 } \right)}}{{\Delta ^2  - \Omega ^2  + {\gamma  \mathord{\left/
 {\vphantom {\gamma  {\kappa \beta }}} \right.
 \kern-\nulldelimiterspace} {\kappa \beta }}}}, \\
 \frac{\alpha }{\mu }\Xi ^2  = \frac{{2\alpha }}{{3\mu }}\Delta ^2  + 1 - \varsigma  + c, \\
 p\left( \omega  \right) \approx \frac{{6\pi \beta ^2 }}{{\alpha \gamma }}\frac{{\Delta ^2  - \omega ^2  + {\gamma  \mathord{\left/
 {\vphantom {\gamma  {\kappa \beta }}} \right.
 \kern-\nulldelimiterspace} {\kappa \beta }}}}{{\Xi ^2  - \omega ^2 }}{\mathop{\rm H}\nolimits} \left( {\Delta ^2  - \omega ^2 } \right), \\
 \end{gathered}
 \end{eqnarray}
\noindent where $\varsigma\equiv \sigma/\mu$, $
c \equiv {{\alpha \gamma } \mathord{\left/
 {\vphantom {{\alpha \gamma } {\beta \mu \kappa }}} \right.
 \kern-\nulldelimiterspace} {\beta \mu \kappa }}
$
 and $
\Upsilon  \equiv \left( {2 - c} \right)^2  - 4\varsigma \left( {2 - \varsigma  + c} \right)
$. 

It should be noted, that our technique based on the adiabatic approximation and the regularization of $d\Omega/dt$ \cite{kalash1,kalash2,kalash4,kalash5,kalash6,kalash7,kalash8,kalash11} is analogous to that of Ref. \cite{abl}. However, an approximate integration in the spectral domain allows us the further reduction of parametric space dimension and the construction of physically meaningful master diagrams, which make the CDS properties to be easily traceable.

The CDS energy is

\begin{eqnarray} \label{energy}
\begin{gathered}
  E \equiv \int\limits_{ - \infty }^\infty  {P\left( t \right)dt}
  \approx \int\limits_{ - \Delta }^\Delta  {p\left( \omega  \right)} \frac{{d\omega }}
{{2\pi }} =  \hfill \\
  \frac{{6\beta ^2 \Delta }}
{{\alpha \gamma }}\left[ {1 - \frac{{\left( {\Xi ^2  - \Delta ^2  -
{\gamma  \mathord{\left/
 {\vphantom {\gamma  {\beta \varsigma }}} \right.
 \kern-\nulldelimiterspace} {\beta \varsigma }}} \right)\operatorname{arctanh} \left( {\frac{\Delta }
{\Xi }} \right)}}
{{\Delta \Xi }}} \right]. \hfill \\
\end{gathered}
\end{eqnarray}

The normalizations of the previous subsection are useful in this case with the replacements $\kappa \rightarrow \mu \kappa$, $\zeta \rightarrow \kappa$. Then the dimensionless CDS energy is

\begin{equation}\label{en}
    E = \frac{{6\Delta }}{{c^2 }}\left[ {1 - \frac{{\left( {\Xi ^2  - \Delta ^2  - c} \right){\mathop{\rm arctanh}\nolimits} \left( {{\Delta  \mathord{\left/
 {\vphantom {\Delta  \Xi }} \right.
 \kern-\nulldelimiterspace} \Xi }} \right)}}{{\Delta \Xi }}} \right].
\end{equation}

\begin{figure}[h!]
\includegraphics[width=7.3cm]{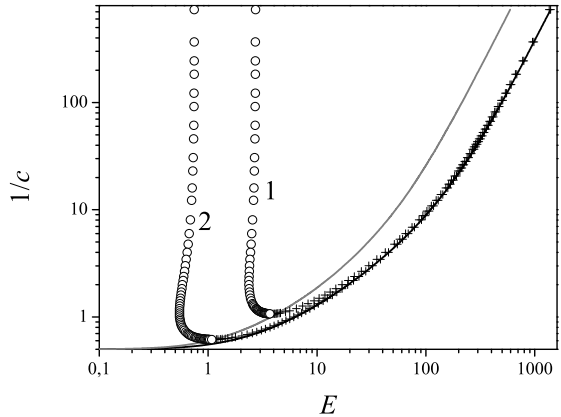}
\caption{\label{fig5} The master diagram for the CDS of Eqs. (\ref{GL},\ref{sesam}). There exists no
CDS below the black solid line (here $\sigma<$0, i.e. the vacuum of (\ref{GL}) is unstable). Black solid curve corresponds to $\varsigma=$0. Gray curve divides the regions, where the $+$ and $-$
branches of (\ref{sesam2}) exist. Crosses (circles) correspond to the $+$($-$) branches for
$\varsigma$=0.1 (1) and $\varsigma$=0.01 (2).}
\end{figure}

The master diagram following from Eq. (\ref{en}) is presented in Fig. \ref{fig5}. The CDS is two-parametric in this case. The structure of diagram is similar to
that described in the previous subsection, but there are two important differences: i) there is no maximum value of $1/c$-parameter with growth of $E$, ii) $-$ branch is not energy-scalable within all range of existence.

Eq. (\ref{en}) allows obtaining the practical important expression for the threshold energy of $+$branch of CDS providing its stabilization against an excitation of the vacuum of Eq. (\ref{GL}). Such an expression correspond to $\varsigma=$0 in Eq. (\ref{sesam2}) with $+$ sign before $\sqrt{\Upsilon}$ (black solid curve in Fig. \ref{fig5}):

\begin{equation}\label{threshold}
    E_{th}  = \frac{3}{{c^2 }}\left[ {\sqrt {\frac{{6\left( {2 - c} \right)}}{c}}  - \frac{{3\left( {2 + c\left( {c - 2} \right)} \right)}}{{\sqrt {1 - c\left( {c - 3} \right)} }}{\mathop{\rm arctanh}\nolimits} \left( {\frac{{\sqrt {6c\left( {2 - c} \right)} }}{{2\sqrt {1 - c\left( {c - 3} \right)} }}} \right)} \right].
\end{equation}

Eq. (\ref{threshold}) allows estimating the threshold GDD from the desired level of the CDS energy and oscillator parameters. Table \ref{table} based on Eq. (\ref{threshold}) presents the threshold values of GDD for a Yb:YAG thin-disk oscillator with 1 MHz repetition rate, 50$\div$100 W average output power, and the different semiconductor saturable absorbers (courtesy of R.Graf, Max-Planck-Institut f\"{u}r Quantenoptik, Garching  and Dr.A.Apolonski, Ludwig-Maximilians-Universit\"{a}t M\"{u}nchen, Garching).

\begin{table}[h]
\begin{center}
\caption{\label{table}Parameters (courtesy of R.Graf and Dr.A.Apolonski) of a Yb:YAG thin-disk oscillator with 10\% output coupler: $\alpha= $3.6$\times 10^{4}$
fs$^2$, $T_r$=1 ps. $\gamma=$0.22 GW$^{-1}$ corresponds to an airless resonator with 4 transits through an active medium. Beam sizes on an active medium and a semiconductor saturable absorber are 2.5 mm and 0.8 mm, respectively.
The spectrum is centered at $\approx$1 $\mu$m.}
\begin{tabular}{|l|l|l|l|l|l|}
\hline
Set & $\beta$, ps$^2$& 2$\Delta$, ps$^{-1}$ &$E_{out}$, $\mu$J & $E_s$, $\mu$J/cm$^2$& $\mu$\\
\hline
1 & 0.024& 0.6& 50 & 90& 0.002\\
 & 0.03& 0.6& 100 & 90& 0.002\\
\hline
2 & 0.014&0.8 & 50 & 35& 0.0042\\
 & 0.018&0.8 & 100 & 35& 0.0042\\
\hline
3 & 0.0085& 1.7& 50 & 61& 0.018\\
 & 0.011& 1.7& 100 & 61& 0.018\\
\hline
\end{tabular}
\end{center}
\end{table}

One can see from Table \ref{table}, that the CDS provides the output pulse energy levels of 50$\div$100 $\mu$J for the $\beta$-levels, which are achievable by the multipass chirped-mirrors delay lines. The minimum $\beta$ and the maximum spectral widths are provided by the largest modulation depth $\mu$ (Set 3 in Table \ref{table}). Insertion of an additional SPM source (e.g., a fused silica plate) will result in the $\beta$ growth, but also it will increase the CDS spectral width and, thereby, reduce the pulse width after compression.

\section{Conclusion}

The theory of chirped dissipative solitons is exposed on the basis of the approximate integration method of the nonlinear complex Ginzburg-Landau equation. Two main types of nonlinearity are considered: cubic-quintic and perfectly saturable ones, but the theory is not confined to a functional form of nonlinearity because only its instantaneous response is required. The main assumption allowing the approximate solution is dominance of nondissipative factors over the dissipative ones.

The advantage of the obtained CDS solutions, which have truncated spectra with concave, convex and concave-convex vertexes, is their reduced parametric dimension. This makes the CDS to be easily traceable within an extremely broad range of parameters.

An application of the theory under consideration to an energy-scalable mode-locked oscillator is presented. It is pointed, that scaling properties of CDS allow reaching the sub-mJ energies of sub-picosecond pulses directly from a thin-disk solid-state oscillator operating at the MHz repetition rate. This promise a breakdown in further high-intensity laser experiments.

This work was supported by the Austrian
Fonds zur F\"{o}rderung der wissenschaftlichen Forschung
(FWF Project No. P20293).

\label{last}

\end{document}